%% file: main.tex
\documentclass{article}
\usepackage{spconf,amsmath,graphicx,hyperref}
\usepackage{booktabs}
\usepackage{amssymb}
\usepackage[table]{xcolor}
\usepackage{xspace}
\usepackage{caption}
\usepackage{microtype}
\usepackage{enumitem}

\definecolor{realgray}{RGB}{235,235,235}
\definecolor{oursblue}{RGB}{225,240,250}

\newcommand{\ourmodel}{\textsc{Define}\xspace}

\title{DEFINE: Exemplar-Guided Accent Control for Zero-Shot TTS}
\name{Ambuj Mehrish$^{1}$, Abhinaba Roy$^{4}$, Alex Ivanov$^{2}$, Tawsif Ahmed$^{3}$, Dorien Herremans$^{4}$}
\address{
$^{1}$Ca' Foscari University of Venice \quad
$^{2}$Kandinsky Lab \quad $^{3}$Sleeping AI \\
$^{4}$Singapore University of Technology and Design
}
\begin{document}
\ninept
\maketitle
\begin{abstract}
Zero-shot text-to-speech (TTS) can reproduce an unseen speaker from a short reference recording, but typically entangles speaker identity and accent within the same reference. We introduce \ourmodel{}, an end-to-end framework that decouples these factors by conditioning speaker identity and target accent on separate audio exemplars. A single inference-time guidance weight continuously controls accent strength without retraining. Built on F5-TTS with parameter-efficient LoRA adaptation, \ourmodel{} maps short accent exemplars into a conditioning space using an exemplar encoder supervised through learned accent prototypes, requiring neither accent labels at inference time nor post-synthesis waveform conversion. On seen accents, increasing accent guidance improves accent-probe accuracy from 6.5\% to 19.6\%. More importantly, a single \ourmodel{} model generalizes accent control beyond its training accent set: on seen and out-of-domain accents, though not on held-out accents, it matches the accent transfer performance of a two-model TTS--voice-conversion cascade while achieving higher speaker similarity and comparable predicted speech quality. These results demonstrate that speaker identity and accent can be independently controlled from audio exemplars within a single zero-shot TTS model, including for accents unseen during training.
\footnote{\href{https://github.com/AMAAI-Lab/define}{https://github.com/AMAAI-Lab/define}}
\end{abstract}
\section{Introduction}
\label{sec:intro}
Recent zero-shot TTS models can reproduce the voice of an unseen speaker from only a few seconds of reference speech~\cite{chen2025f5,mehrish2023review,eskimez2024e2}. However, the reference carries more than speaker identity: it also carries the speaker's accent. There is no independent mechanism for requesting, for example, \emph{the reference speaker's voice with a Scottish accent}. This coupling is reinforced by training data in which each speaker is observed with a single native accent. The training objective therefore provides little incentive to represent them as independently controllable factors.

Previous work has approached this problem in two main ways. The first introduces accent as an explicit conditioning variable, represented by an accent label or an embedding learned from labeled speech~\cite{melechovsky2024dart,melechovsky2023learning}. A second approach performs accent conversion after synthesis, passing generated speech through a separate accent or voice-conversion model~\cite{liu2024zero}. More importantly, neither formulation naturally provides a unified mechanism in which \emph{voice} and \emph{accent} are independently specified by reference examples, while accent strength remains controllable after training.

We introduce \ourmodel{} (Disentangled Exemplar Framework for Identity and Novel-accent Expression), a modular approach that explicitly separates these two sources of conditioning. The conventional reference audio specifies \emph{who} should speak, while a separate short accent exemplar specifies \emph{how} the generated speech should be accented. The accent exemplar is encoded directly into a conditioning representation and injected into the TTS model, avoiding a separate waveform-conversion stage. A key feature of \ourmodel{} is that at inference time, a single guidance weight $w$ controls the contribution of accent conditioning. Setting $w{=}0$ removes the contribution of accent-conditioning, while increasing $w$ progressively strengthens the influence of the target accent. 

However, learning a useful global accent representation from the flow-matching objective alone provides only indirect supervision to the accent encoder. We address this with \emph{prototype anchoring}. An exemplar encoder maps the accent clip to a conditioning vector, while a small learned table maintains one prototype for each training accent. During training, these prototypes provide stable targets that organize the accent representation space and supervise the exemplar encoder. The resulting representation is injected alongside the timestep and text conditioning of an F5-TTS~\cite{chen2025f5} backbone. We keep the pretrained backbone frozen and perform parameter-efficient adaptation using LoRA~\cite{hu2021lora}. Our main contributions are:

\begin{itemize}[left=0pt]
\setlength{\itemsep}{0pt}
  \setlength{\parsep}{0pt} 
\item We introduce \ourmodel{}, a zero-shot TTS framework that independently specifies speaker identity and target accent from separate audio examples, without requiring an accent label at inference time. 

\item We propose \emph{prototype anchoring}, in which learned per-accent prototypes provide direct supervision for an exemplar encoder, enabling short accent examples to be mapped to a stable conditioning space despite the weak accent supervision provided by the flow-matching objective alone.

\item We introduce continuous inference-time accent control through a single guidance weight $w$, allowing accent strength to be varied after training.
\end{itemize}

    

\begin{figure*}
    \centering
    \includegraphics[width=0.80\linewidth]{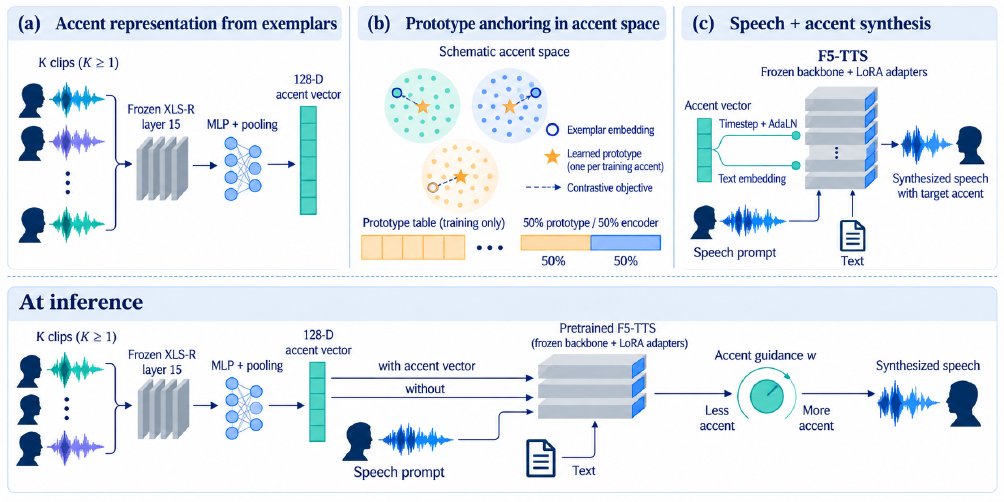}
    \caption{\ourmodel{} overview. (a) Accent exemplars are encoded by a frozen XLS-R and pooled into one accent vector; (b) a learned per-accent prototype table supervises the encoder during training. The accent vector shifts the timestep and text embeddings of a frozen F5-TTS backbone adapted with LoRA. At inference the prototypes are discarded and a single weight $w$ sets accent strength. Snowflake: frozen, flame: trained.}
    \label{fig:method}
\end{figure*}
\section{Related Work}
\label{sec:related}

Our work relates to three lines of research: zero-shot text-to-speech, explicit accent conditioning, and post-hoc accent conversion. Zero-shot TTS synthesizes speech in the voice of an unseen speaker from a short reference utterance. Recent approaches use discrete speech representations~\cite{wang2023neural} or flow matching over masked acoustic features~\cite{le2023voicebox,eskimez2024e2,chen2025f5}, while earlier systems rely on explicit speaker encoders~\cite{casanova2022yourtts}. In these formulations, the reference jointly specifies speaker identity and speaking characteristics, implicitly coupling accent with voice. Other work introduces accent as an explicit conditioning variable through labels or learned representations~\cite{melechovsky2023learning,melechovsky2024dart,cong2023genertts}, with some methods additionally controlling accent strength~\cite{liu2024controllable}. These approaches provide direct control but are generally restricted to accents represented during training, requiring labelled data and model adaptation for new accents. Alternatively, accent conversion modifies synthesized or recorded speech~\cite{zhao2019foreign,ding2022accentron,jia2024convert,jin2023voice}, often using a separate voice-conversion model~\cite{liu2024zero}, introducing an additional transformation stage in which speaker characteristics must be preserved. 

\section{Method}
\label{sec:method}
\subsection{Backbone and accent injection}
 We build on F5-TTS~\cite{chen2025f5} (Figure~\ref{fig:method}), a non-autoregressive text-to-speech model trained with conditional flow matching~\cite{lipman2022flow} on mel spectrograms. Let $x_1\in\mathbb{R}^{T\times F}$ denote the target mel spectrogram, $x_0\sim\mathcal{N}(0,I)$, and $t\sim\mathcal{U}(0,1)$. Given $x_t=(1-t)x_0+tx_1$, the network $v_\theta$ learns the conditional vector field by minimizing:
\begin{equation}
\begin{split}
\mathcal{L}_{\mathrm{fm}}
&= \mathbb{E}_{t,x_0,x_1}
   \left[\left\|m\odot\Delta v_\theta\right\|_2^2\right],\\
\Delta v_\theta
&= v_\theta(x_t,t,c,y,a)-(x_1-x_0).
\end{split}
\label{eq:fm}
\end{equation}
here, $y$ is the text, $c$ is the reference audio that carries speaker identity through in-context infilling, $m$ is a binary target-span mask, and $a\in\mathbb{R}^{d}$ is an accent embedding with $d=128$. The backbone architecture is unchanged. Rank-16 LoRA adapters~\cite{hu2021lora} are applied to the attention query and value projections, feed-forward layers, and adaLN modulation, resulting in $6.3$M trainable parameters out of $343$M. The accent vector enters at two places (Fig.~\ref{fig:method}). Let $\rho(u)=\sqrt{\frac{1}{n}\sum_{i=1}^{n}u_i^2}$ denote the root-mean-square (RMS) of an $n$-dim vector, and $\mathrm{LN}$ layer normalization.
The timestep and text embeddings are shifted as follows:
\begin{equation}
\begin{split}
e_t &\leftarrow e_t+\gamma_t\,\rho(e_t)\,\mathrm{LN}(W_ta),\\
e_y &\leftarrow e_y+\gamma_y\,\rho(e_y)\,\mathrm{LN}(W_ya).
\end{split}
\label{eq:inject}
\end{equation}
RMS scaling matches each shift to the size of the embedding it modifies; without it the added term is about 100 times smaller than the timestep embedding and is ignored during training. Since $e_t$ controls adaLN, accent conditioning reaches every transformer block.
$W_t$ and $W_y$ are bias-free and $\mathrm{LN}$ is non-affine, so $a{=}0$ leaves
both embeddings unchanged. During training, $a$ is set to $0$ with $p_{\mathrm{drop}}=0.15$ to train the accent-free inference branch.

\subsection{Exemplar encoder and prototype anchoring}

Let $\mathcal{X}=\{u_1,\dots,u_K\}$ be $K\!\ge\!1$ short clips of the desired accent from speakers absent from training (Fig.~\ref{fig:method}a). Training uses $K{=}1$; at inference we pool over the provided clips, with $K{=}1$ and $K{=}3$ giving similar accuracy (Sec.~\ref{sec:results}). Each clip is cropped to speech regions and encoded by a frozen XLS-R~\cite{babu2021xls} model, read at layer $\ell{=}15$ as selected by a layer-wise accent probe. Let $h_k\in\mathbb{R}^{T_k\times 1024}$ be the frame features of clip $k$, let $\overline{h_k}$ be their masked mean over frames, and write $\nu(u)=u/\|u\|_2$. An MLP $g_\phi$ maps each pooled clip to $\mathbb{R}^{d}$, and the clip embeddings are averaged:
\begin{equation}
\bar a_k=\nu\big(g_\phi(\overline{h_k})\big),
\qquad
a_{\mathrm{enc}}=\nu\Big(\frac{1}{K}\sum_{k=1}^{K}\bar a_k\Big).
\label{eq:enc}
\end{equation}
Training $g_\phi$ with \eqref{eq:fm} alone provides only weak and indirect supervision for global conditioning, resulting in embeddings that do not effectively guide the decoder (see Table~\ref{tab:ablation}). An auxiliary \emph{prototype table} is introduced, $P\in\mathbb{R}^{C\times d}$
with one row per training accent ($C{=}23$; Fig.~\ref{fig:method}b), normalised as $p_c = P_c/\|P_c\|_2$.
The encoder is aligned with these prototypes using cosine and contrastive terms:
\begin{equation}
\mathcal{L}_{\mathrm{proto}}
=1-a_{\mathrm{enc}}^\top\mathrm{sg}[p_c]
-\lambda_{\mathrm{ce}}\log
\frac{\exp(a_{\mathrm{enc}}^\top p_c/\tau)}
{\sum_{j=1}^{C}\exp(a_{\mathrm{enc}}^\top p_j/\tau)}.
\label{eq:proto}
\end{equation}
Here, $\tau=0.1$ and $\mathrm{sg}$ denotes stop-gradient. To reduce dependence on exact prototype lookup, the decoder is conditioned during training on a Bernoulli mixture: $a = b\,\mathrm{sg}[a_{\mathrm{enc}}]+(1-b)\,p_c$, where $b \sim \mathrm{Bernoulli}(\pi)$ and $\pi=0.5$. The stop-gradient operation prevents $\mathcal{L}_{\mathrm{fm}}$ from updating $g_\phi$, ensuring that the prototypes learn decoder-compatible conditioning while the encoder learns to match them. At inference, $P$ is discarded and only \eqref{eq:enc} is used (see Fig.~\ref{fig:method}, bottom), enabling representation of unseen accents directly from their exemplars. Two auxiliary losses shape the embedding space. Supervised contrastive loss~\cite{khosla2020supervised}
$\mathcal{L}{\mathrm{con}}$ brings clips with the same accent label together. A speaker classifier with a gradient reversal layer~\cite{ganin2015unsupervised} supplies $\mathcal{L}_{\mathrm{spk}}$, discouraging speaker information in $a_{\mathrm{enc}}$. The full objective is
\begin{equation}
\begin{split}
\mathcal{L}
= \mathcal{L}_{\mathrm{fm}}
 + \lambda_p\mathcal{L}_{\mathrm{proto}} +\lambda_c\mathcal{L}_{\mathrm{con}}
 + \lambda_s\mathcal{L}_{\mathrm{spk}}.
\end{split}
\label{eq:total}
\end{equation}

\subsection{Accent guidance at inference}

Accent strength is adjustable in a continuous manner during inference.
Let $v_\varnothing=v_\theta(x_t,t,c,y,0)$ denote the
accent-free prediction, $v_{\mathrm{u}}$ the unconditional
prediction for classifier-free guidance~\cite{ho2022classifier},
and $v_a=v_\theta(x_t,t,c,y,a)$. At each solver step, we use
\begin{equation}
\hat{v} = v_\varnothing + s\,(v_\varnothing - v_{\mathrm{u}}) + w\,(v_{a} - v_\varnothing),
\label{eq:guide}
\end{equation}
with $s=2$; here $v_u$ drops both the text and the audio prompt. If $w=0$, the adapted model is used without accent conditioning. Increasing $w$ makes the accent contribution stronger, allowing a balance between speaker similarity and accent strength (see Table~\ref{tab:weights}). 



\section{Experimental Setup and Results}
\label{sec:exp}
\input{Table/table_foldA}
\input{Table/table_weights}
\noindent\textbf{Data.} Training data comes from two sources. The English subset of Common Voice~\cite{ardila2020common} supplies naturally accented speech from many speakers per accent; after removing one low-resource accent we retain $15$, holding out Australian, Irish and Filipino for unseen-accent evaluation and reserving Welsh and West Indian as out-of-domain conditions. Training also covers 13 further accent classes that are never evaluated, mostly L2-English varieties, so the prototype table has $C{=}23$ rows: the 10 trained accents of the evaluation taxonomy plus these 13. Held-out accents appear in the probe's label set; out-of-domain accents are
absent from training in either fold. Speaker prompts come from an in-house set of studio-quality recordings synthesised with a commercial TTS system, annotated per speaker for accent and gender under the same taxonomy, and kept only when word-aligned prompt--target splits meet duration, transcription-confidence, speech-presence and script quality criteria; these prompts give the highest speaker similarity in Table~\ref{tab:main}.

Because Common Voice couples the two factors, we use Seed-VC~\cite{liu2024zero} to transfer Common Voice utterances into the timbre of studio speakers with different accents, yielding $94.7$k deliberately mismatched prompt--target pairs alongside $137$k real recordings. Held-out and out-of-domain accents are excluded
from training, and exemplar speakers are disjoint from training speakers.

\noindent\textbf{Evaluation.}
All systems are evaluated on the same set of 1,764 items: 1,134 seen, 378 held-out, and 252 out-of-domain; each a voice prompt, at least one exemplar clip from a different target accent, and unseen text. Accent accuracy (ACC) uses a 15-way logistic-regression probe on layer-15 XLS-R features~\cite{babu2021xls} (chance${=}0.067$)\footnote{The probe shares its feature extractor with the exemplar encoder, but does not appear to favour it: the prototype-lookup path uses no XLS-R at inference yet scores highest ($0.354$ vs.\ $0.196$), and the cascade uses none
anywhere yet ties DEFINE ($0.188$ vs.\ $0.196$).}. Speaker similarity (SPK) is the cosine similarity between ECAPA-TDNN embeddings~\cite{desplanques2020ecapa} of the output and the prompt. Word error rate (WER) comes from Whisper large-v3~\cite{radford2023robust}, and predicted quality from UTMOS~\cite{saeki2022utmos}, with 95\% bootstrap intervals calculated over items. Baselines include F5-TTS~\cite{chen2025f5}, \ourmodel{} with accent guidance disabled ($w{=}0$; LoRA remains active), and an F5-TTS--Seed-VC cascade. Real target-accent recordings serve as a reference. Additionally, we include a non-deployable prototype-lookup oracle that receives the accent label and retrieves its row from $P$.
\subsection{Results}
\label{sec:results}
\noindent\textbf{Accent control and guidance.}
For seen accents, \ourmodel{} raises ACC from 0.065 with guidance disabled to 0.196 at $w{=}15$, matching the Seed-VC cascade (0.188) with 0.024 higher SPK and equivalent predicted quality (Table~\ref{tab:main}). Unlike waveform conversion, \ourmodel{} modulates accent within the synthesizer instead of re-rendering the generated voice through a separate converter, so identity is better preserved. Each solver step uses one fused classifier-free call and one accent-conditioned call: three sequence evaluations against two for F5-TTS, whereas the cascade adds a full conversion pass over the waveform. On accents absent from all training data \ourmodel{} is statistically
indistinguishable from the cascade ($0.087$ vs.\ $0.083$) and improves over the same model with the accent disabled by $+0.040$ ACC (95\% CI $[+0.004,+0.079]$, paired over items) and over F5-TTS by $+0.064$ ($[+0.032,+0.099]$), though the cascade remains stronger on held-out accents ($0.175$ vs.\ $0.127$). The sweep in Table~\ref{tab:weights} reveals a smooth trade-off: SPK falls monotonically with $w$, so a single post-training scalar picks the desired operating point. Exemplar-based ACC peaks at $w{=}19$ for both seen (0.214) and OOD accents (0.111), whereas prototype lookup continues to improve to 0.385 at $w{=}23$, which points to  accent inference from exemplars, not the conditioning pathway, as the  bottleneck at high guidance.
\input{Table/ablation}

\noindent\textbf{Ablated variants.} Beyond removing individual terms of \eqref{eq:total}, we compare three alternatives. \emph{Cross-attention conditioning} replaces the additive injection of \eqref{eq:inject} with a zero-initialised cross-attention layer inserted after every fourth backbone block, attending from the hidden states to the XLS-R exemplar frames rather than to a single pooled vector. The \emph{encoder-consistency loss} adds a round-trip term: the predicted mel span is decoded with the vocoder, re-encoded with the exemplar encoder, and penalised by $1-\cos(a_{\mathrm{gen}}, \mathrm{sg}[a])$, so that generated audio carries back the accent vector it was conditioned on. Finally we raise the Bernoulli mixing rate to $\pi{=}0.7$ and repeat the converted rows twice in the training pool. The gradient variant detaches the accent input entirely, so $\mathcal{L}_{fm}$
no longer updates the prototype table either; the encoder is stop-gradiented in
all variants.

\noindent\textbf{Ablations and bottleneck analysis.} Without prototype anchoring, no guidance weight improves on switching the accent off ($w^\ast{=}0$, ACC 0.079): an encoder trained from $\mathcal{L}_{\mathrm{fm}}$ alone does not steer the decoder. The contrastive prototype term adds a further 0.053 (Table~\ref{tab:ablation}). Variants must be compared at matched quality: pushed to $w{=}23$,
cross-attention reaches 0.175 ACC but collapses to UTMOS 1.52 and SPK 0.22 --- unintelligible audio that still fires the probe, which is why its usable
weight in Table~\ref{tab:ablation} is $0$. Mean-pooling three exemplars instead of one changes ACC by $+0.000$ (95\% CI $\pm0.058$, seen) and $+0.003$ (CI $[-0.026,+0.032]$, held-out), and adds 0.019 SPK. The limit is the encoder itself: prototype lookup reaches 0.354 ACC on seen accents against 0.196 from exemplars, and its nearest-prototype accuracy falls from 0.873 on training speakers to 0.540 on unseen ones (chance $1/23 = 0.043$). The decoder follows a well-placed accent vector; speaker overfitting in the encoder, not conditioning capacity or guidance, is the bottleneck.

Seed-VC fulfills two primary roles: during augmentation, it transfers timbre while the Common Voice utterance supplies the accent, whereas the baseline model is required to transfer the accent. The observed parity on seen accents at higher speaker similarity cannot be explained by exposure to Seed-VC outputs. This is not what one would expect from a model distilling the cascade; however, \ourmodel{} maintains a speaker similarity of $0.624$ to $0.654$, compared to $0.600$ to $0.603$ for the baseline. Held-out and out-of-domain accents are excluded from the converted pool, ensuring that the out-of-domain result ($0.087$ vs.\ $0.083$) does not utilize any Seed-VC-generated training data.




\begin{figure}
    \centering
    \includegraphics[width=0.9\columnwidth]{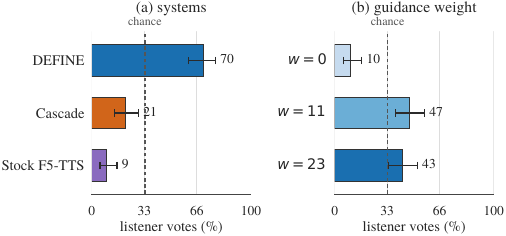}
 \caption{Listening test on seen accents. All \ourmodel{} samples use prototype-lookup, i.e.\ accents given as a label, at $w{=}11$, where the accent change is clearly perceptible. (a) \ourmodel against F5-TTS and the cascade; (b) the same model at three guidance weights. 11 listeners, 10 trials per part, giving 110 votes per part. Bars show the share of votes each option received, with 95\% Wilson intervals.}
    \label{fig:listening}
\end{figure}

\subsection{Listening test}
\label{sec:listening}
To determine how the accent changes are perceived by the human ear, we ran a listening test on the seen-accent set. Samples use the prototype-lookup configuration, which fixes the conditioning path so the test isolates accent control from encoder quality: decoder, injection and guidance are identical to the exemplar system. 11 listeners completed two parts of ten trials each (four England targets, three India, three US). Each trial gave a reference clip of the target accent, a clip of the reference voice, and three synthesised samples in random order, and the listener picked the sample closest to the target accent. Each part therefore gives 110 votes against a chance rate of $0.333$. Fig.~\ref{fig:listening} summarises both parts.

\noindent\textbf{Systems.} The first part puts \ourmodel{} at $w{=}11$ against F5-TTS and the F5-TTS--Seed-VC cascade. \ourmodel{} took 77 of the 110 votes ($0.70$, 95\% CI $[0.61,0.78]$), the cascade 23 ($0.21$) and F5-TTS 10 ($0.09$), \ourmodel{} was the most-picked system for 9 of 11 listeners (binomial test against
chance, $p=0.0014$). Listeners thus favour the single-pass output over the two-model cascade by a wide margin. 

\noindent\textbf{Guidance weight.} The second part compares $w{=}0$, $w{=}11$ and $w{=}23$ from the same model, with 11, 52 and 47 votes respectively. Guidance off falls far below chance ($p<10^{-7}$), so the accent listeners hear comes from the guidance term and not from the LoRA adaptation on its own. The two active settings are not separable ($p{=}0.69$). Since speaker similarity keeps falling as $w$ grows (Table~\ref{tab:weights}), a moderate $w$ is the better operating point.

\section{Conclusion}
\label{sec:conclusion}
We introduced \ourmodel{}, an exemplar-based method for zero-shot TTS that separates speaker identity, defined by a reference prompt, from accent, specified by short exemplar clips. A post-training scalar controls accent strength without retraining. Prototype anchoring aligns the exemplar encoder with per-accent prototypes learned through flow matching, providing direct supervision that improves accent control while maintaining baseline intelligibility and predicted quality at practical operating points. With a single inference pass and no accent labels, \ourmodel{} matches the conversion cascade on seen accents, is statistically indistinguishable from it on out-of-domain accents, and better preserves speaker identity.

\section{Acknowledgements}
This work has received support from MOE under grant number MOE-T2EP20124-0014, and SUTD GAP-052 project. We acknowledge the EuroHPC Joint Undertaking for access to LEONARDO at CINECA, Italy, through the EuroHPC AI Factories call “AI for Science and Collaborative EU Projects” (Proposal No. EHPC-AIF-2026SC01-041).


\bibliographystyle{IEEEbib}
\bibliography{strings,refs}

\end{document}

%% file: Table/table_foldA.tex
\begin{table}[t]
\centering
\caption{Accent control with 95\% bootstrap intervals over items. \ourmodel is reported at $w{=}15$, selected on validation items disjoint from the evaluation grid; it infers the accent from exemplar clips and uses no accent
label. $^\dagger$\emph{Prototype lookup} is instead given the accent label and reads the corresponding row of the learned prototype table; it therefore cannot address accents outside the training set; it is also shown at $w{=}11$, the operating point
of the listening study. The reference row is real speech by a different speaker reading different text,
so SPK and WER against the prompt and target sentence do not apply (\texttt{--}).}
\label{tab:main}

\setlength{\tabcolsep}{3pt}
\resizebox{0.9\columnwidth}{!}{%
\begin{tabular}{l r@{\hspace{4pt}}l r r r}
\toprule
System
& \multicolumn{2}{c}{ACC $\uparrow$ (95\% CI)}
& SPK $\uparrow$
& WER $\downarrow$
& UTMOS $\uparrow$ \\
\midrule

\multicolumn{6}{l}{\emph{Seen accents}, $n=1134$ items} \\
\midrule
\rowcolor{realgray}
Real speech (reference)
& 0.436 & [0.406, 0.464] & \texttt{--} & \texttt{--} & 3.08 \\
F5-TTS (stock)
& 0.079 & [0.063, 0.096] & 0.732 & 0.071 & 3.99 \\
Cascade F5-TTS $\rightarrow$ Seed-VC
& 0.188 & [0.166, 0.211] & 0.600 & 0.056 & 3.92 \\
\rowcolor{oursblue}
\ourmodel, accent off
& 0.065 & [0.051, 0.079] & 0.694 & 0.063 & 4.06 \\
\rowcolor{oursblue}
\ourmodel (exemplar)
& 0.196 & [0.173, 0.219] & 0.624 & 0.067 & 3.93 \\
\rowcolor{oursblue}
\quad \emph{-- prototype lookup}$^\dagger$, $w{=}11$
& 0.326 & [0.300, 0.354] & 0.592 & 0.075 & 3.88 \\
\rowcolor{oursblue}
\quad \emph{-- prototype lookup}$^\dagger$, $w{=}15$
& 0.354 & [0.327, 0.381] & 0.562 & 0.089 & 3.77 \\

\midrule
\multicolumn{6}{l}{\emph{Held-out accents}, $n=378$ items} \\
\midrule
\rowcolor{realgray}
Real speech (reference)
& 0.455 & [0.402, 0.505] & \texttt{--} & \texttt{--} & 3.18 \\
F5-TTS (stock)
& 0.061 & [0.037, 0.087] & 0.737 & 0.073 & 3.99 \\
Cascade F5-TTS $\rightarrow$ Seed-VC
& 0.175 & [0.138, 0.212] & 0.603 & 0.076 & 3.89 \\
\rowcolor{oursblue}
\ourmodel, accent off
& 0.066 & [0.042, 0.093] & 0.698 & 0.057 & 4.05 \\
\rowcolor{oursblue}
\ourmodel (exemplar)
& 0.127 & [0.095, 0.161] & 0.654 & 0.062 & 4.01 \\

\midrule
\multicolumn{6}{l}{\emph{Out-of-domain accents}, $n=252$ items} \\
\midrule
\rowcolor{realgray}
Real speech (reference)
& 0.333 & [0.278, 0.389] & \texttt{--} & \texttt{--} & 3.06 \\
F5-TTS (stock)
& 0.024 & [0.008, 0.044] & 0.728 & 0.080 & 3.99 \\
Cascade F5-TTS $\rightarrow$ Seed-VC
& 0.083 & [0.052, 0.119] & 0.601 & 0.047 & 3.94 \\
\rowcolor{oursblue}
\ourmodel, accent off
& 0.048 & [0.024, 0.075] & 0.692 & 0.076 & 4.06 \\
\rowcolor{oursblue}
\ourmodel (exemplar)
& 0.087 & [0.056, 0.123] & 0.640 & 0.074 & 3.97 \\
\bottomrule
\end{tabular}%
}
\end{table}

%% file: Table/table_weights.tex
\begin{table}[t]
\centering
\small
\caption{Guidance sweep. $w{=}0$ disables the accent term; the LoRA adapters remain active, so this is not identical to F5-TTS; larger
$w$ pushes further toward the accent. ACC$_{\mathrm{ex}}$ and SPK are
exemplar-driven; ACC$_{\mathrm{lk}}$ is the
prototype-lookup bound and exists for seen accents only.}
\label{tab:weights}
\resizebox{0.8\columnwidth}{!}{%
\begin{tabular}{lcccccc}
\toprule
& \multicolumn{3}{c}{Seen accents} & \multicolumn{2}{c}{Held-out accents} & Out-of-domain \\
\cmidrule(lr){2-4} \cmidrule(lr){5-6} \cmidrule(lr){7-7}
$w$ & ACC$_{\mathrm{ex}}$ & SPK & ACC$_{\mathrm{lk}}$ & ACC$_{\mathrm{ex}}$ & SPK & ACC$_{\mathrm{ex}}$ \\
\midrule
0 & 0.065 & 0.694 & 0.065 & 0.066 & 0.698 & 0.048 \\
7 & 0.156 & 0.667 & 0.272 & 0.079 & 0.683 & 0.048 \\
9 & 0.168 & 0.655 & 0.311 & 0.098 & 0.675 & 0.056 \\
11 & 0.185 & 0.644 & 0.326 & 0.108 & 0.669 & 0.063 \\
15 & 0.196 & 0.624 & 0.354 & 0.127 & 0.654 & 0.087 \\
19 & 0.214 & 0.604 & 0.376 & 0.135 & 0.638 & 0.111 \\
23 & 0.212 & 0.587 & 0.385 & 0.151 & 0.624 & 0.103 \\
\bottomrule
\end{tabular}}
\end{table}

%% file: Table/ablation.tex
\begin{table}[t]
\centering
\small
\caption{Ablation on the 378 held-out-accent set. Each variant is reported at
the largest guidance weight $w^\ast$ keeping UTMOS within $0.15$ and speaker
similarity within $0.10$ of the same variant with the accent switched off,
selected on validation items: variants tolerate guidance very differently, so a
common $w$ would reward one that degrades into noise. A variant whose $w^\ast$
is $0$ gains no accent control at any weight; its high speaker similarity and
predicted quality simply reflect the accent being switched off.}
\label{tab:ablation}
\setlength{\tabcolsep}{4pt}
\resizebox{\columnwidth}{!}{%
\begin{tabular}{lcccc}
\toprule
Variant & $w^\ast$ & ACC & SPK & UTMOS \\
\midrule
\textbf{\ourmodel} (full) & 15 & 0.127 & 0.654 & 4.01 \\
\quad $-$ contrastive prototype term & 9 & 0.074 & 0.662 & 4.04 \\
\quad $-$ prototype anchoring (free encoder) & 0 & 0.079 & 0.711 & 4.03 \\
\quad $-$ flow-matching gradient into the accent path & 0 & 0.074 & 0.717 & 4.01 \\
\quad $+$ encoder-consistency loss & 0 & 0.074 & 0.698 & 4.04 \\
\quad cross-attention conditioning & 0 & 0.085 & 0.693 & 4.01 \\
\quad $\pi{=}0.7$ encoder mixing, $2\times$ augmentation & 15 & 0.103 & 0.651 & 4.01 \\
\bottomrule
\end{tabular}}
\end{table}